\documentclass[useAMS,usenatbib]{mn2e}
\usepackage{graphicx}
\title[The multiplicity of planet host stars]
{The multiplicity of planet host stars \newline --- New low-mass companions to planet host stars}
\author[Mugrauer et al.]{M. Mugrauer$^{1}$\thanks{E-mail: markus@astro.uni-jena.de}, A. Seifahrt$^{1, 2}$,
R. Neuh\"auser$^{1}$
\thanks{Based on observations obtained on La Silla in ESO programs 075.C-0098(A), 077.C-0572(A) and 078.C-0376(A), as well as on Paranal in ESO programs 070.C-0557(A), 076.C-0057(A) and 078.C-0376(B).}\\
$^{1}$Astrophysikalisches Institut, Universit\"at Jena, Schillerg\"a{\ss}chen 2-3, 07745 Jena, Germany\\
$^{2}$European Southern Observatory, Karl-Schwarzschild-Str. 2, 85748 Garching, Germany\\}

\begin{document}

\date{Accepted 2007 April 12. Received 2007 April 2; in original form 2007 March 8}

\pagerange{\pageref{firstpage}--\pageref{lastpage}} \pubyear{2007}

\maketitle

\label{firstpage}

\begin{abstract}

We present new results from our ongoing multiplicity study of exoplanet host stars, carried out
with the infrared camera SofI at ESO-NTT. We have identified new low mass companions to the planet
host stars HD\,101930 and HD\,65216. HD\,101930\,AB is a wide binary systems composed of the planet
host star HD\,101930\,A and its companion HD\,101930\,B which is a M0 to M1 dwarf with a mass of
about 0.7\,$M_{\odot}$ separated from the primary by $\sim$73\,arcsec (2200\,AU projected
separation). HD\,65216 forms a hierarchical triple system, with a projected separation of 253\,AU
(angular separation of about 7\,arcsec) between the planet host star HD\,65216\,A and its close
binary companion HD\,65216\,BC, whose two components are separated by only $\sim$0.17\,arcsec
(6\,AU of projected separation). Two VLT-NACO images separated by 3 years confirm that this system
is co-moving to the planet host star. The infrared photometry of HD\,65216\,B and C is consistent
with a M7 to M8 (0.089\,$M_{\odot}$), and a L2 to L3 dwarf (0.078\,$M_{\odot}$), respectively, both
close to the sub-stellar limit. An infrared spectrum with VLT-ISAAC of the pair HD\,65216\,BC, even
though not resolved spatially, confirms this late spectral type. Furthermore, we present H- and
K-band ISAAC infrared spectra of HD\,16141\,B, the recently detected co-moving companion of the
planet host star HD\,16141\,A. The infrared spectroscopy as well as the apparent infrared
photometry of HD\,16141\,B are both fully consistent with a M2 to M3 dwarf located at the distance
of the planet host star.

\end{abstract}


\section{Introduction}

Since the mid nineties of the last century high precision radial-velocity studies revealed more
than 200 exoplanet candidates around mostly solar-like stars. These planet host stars are located
in the solar neighborhood and are mostly isolated single stars. However, already among the first
reported planet detections, three planets were found to orbit the brighter component of binary
systems, namely 55\,Cnc\,AB, $\upsilon$\,And\,AB and $\tau$\,Boo\,AB (Butler et al. 1997).

Since that time more and more planet host multiple star systems were identified, most of them being
found in multiplicity studies of the planet host stars. These studies are carried out with seeing
limited near infrared imaging (see e.g. Mugrauer et al. 2004a\&b, 2005, 2006a) as well as high
contrast diffraction limited AO observations (Patience et al. 2002, Luhman \& Jayawardhana 2002,
Chauvin et al. 2006, and most recently Neuh\"auser et al. 2007). In addition, data from visible and
infrared all sky surveys like POSS or 2MASS are used to identify new companions of planet host
stars (see e.g. Bakos et al. 2006 or Raghavan et al. 2006). Because of all these efforts, more than
30 planet host multiple star systems are known today, suggesting that the multiplicity of planet
host stars is at least 20\,\%.

Most of the detected stellar companions to planet host stars are low-mass main sequence stars. The
projected separations of these companions to the planet host stars range from only a few tens of AU
up to more than 5000\,AU. In a few cases the companions themselves turned out to be close binaries,
i.e. these systems are hierarchical triples (see Mugrauer et al. 2007 for a summary). Not only
main-sequence stars but also white dwarfs were revealed as companions of exoplanet host stars,
suggesting that planetary systems also exist in evolved multiple star systems. Three white dwarfs
were found so far at close (Gl\,86\,B, $\sim$20\,AU, see Mugrauer \& Neuh\"auser 2005, Lagrange et
al. 2006), intermediate (HD\,27442\,B at $\sim$240\,AU, see Chauvin et al. 2006 and Mugrauer et al.
2007) and at wide separations (HD\,147513\,B at $\sim$5400\,AU, see Mayor et al. 2004). Recently,
the first directly imaged substellar companion of an exoplanet host star, the T7.5$\pm$0.5 brown
dwarf HD\,3651\,B, was discovered (see Mugrauer et al. 2006b, Burgasser 2006, Liu et al. 2007, and
Luhman et al. 2007).

In this letter we present new results of our multiplicity study carried out with the ESO-NTT at La
Silla observatory. We have discovered new stellar companions to two planet host stars. We show the
results of our near infrared imaging observations in section\,2. Section\,3 summarizes the infrared
H- and K-band spectroscopic data obtained for the new companions presented here, and also for the
close planet host star companion HD\,16141\,B, for which we have already presented the astrometric
confirmation of companionship in Mugrauer et al. (2005). In the last section we discuss the
properties of the newly found companions.

\section{Observations}

HD\,65216 is a nearby G5 dwarf, located in the southern constellation Carina. Its proper motion and
parallax ($\mu_{\alpha}cos(\delta)=-122.12\pm0.98$\,mas/yr, $\mu_{\delta}=145.90\pm0.64$\,mas/yr,
and $\pi=28.10\pm0.69$\,mas) are both determined by Hipparcos (Perryman et al. 1997), yielding a
distance of $\sim$~36\,pc. According to Santos et al. (2004) HD\,65216 has an effective temperature
of $5666\pm31$\,K and a surface gravity of $log(g)=4.53\pm0.09$\,cm/s$^{2}$, as expected for a mid
G main-sequence star. The same group also determined the mass of the star to be 0.94\,$M_{\sun}$.
Saffe et al. (2005) derived an upper age limit for HD\,65216 to be 10.2\,Gyr, obtained from the
stellar metallicity ($[Fe/H]=-0.12$). A variation of the radial velocity of HD\,65216 with a period
of 613\,day is reported by Mayor et al. (2004), indicating that the star is orbited by a
Jupiter-mass planet ($m \cdot sin(i)=1.21$\,$M_{Jup}$) on an eccentric ($e=0.41$) orbit with a
semi-major axis of $a=1.37$\,AU.

HD\,101930 is a nearby (31\,pc) dwarf of spectral type early K, located in the southern part of the
constellation Centaurus, close to the famous constellation Southern Crux. The proper motion and
parallax of HD\,101930 are both well known from precise Hipparcos measurements
($\mu_{\alpha}cos(\delta)=15.00\pm1.01$\,mas/yr, $\mu_{\delta}=347.49\pm0.80$\,mas/yr, and
$\pi=32.79\pm0.96$\,mas). Lovis et al. (2005) determined its effective temperature to
$5079\pm62$\,K and surface gravity to $log(g)=4.24\pm0.16$\,cm/s$^{2}$). Both values are consistent
with an early K dwarf. The authors also derived the mass of this star to be 0.74\,$M_{\odot}$.
According to Saffe et al. (2005) the age of HD\,101930 ranges between 3.5 and 5.4\,Gyr. HD\,101930
is chromospherically inactive ($logR'_{HK}=-4.99\pm0.02$) but shows a periodical modulation of its
radial velocity (Lovis et al. 2005). This variation of the stellar radial-velocity is induced by a
Saturn-mass exoplanet ($m \cdot sin(i)=0.3$\,$M_{Jup}$), which orbits the star every 70\,days
($a=0.30$\,AU) on a slightly eccentric orbit ($e=0.11$).

We observed HD\,65216 and HD\,101930 for our multiplicity study of planet host stars, using the
infrared camera SofI at the ESO-NTT at La Silla observatory. All observations were obtained in the
H-band with the SofI small-field camera (147\,arcsec$\times$147\,arcsec field of view). By
comparing two images of the planet host stars taken at different observing epochs we can find
co-moving companions of these stars and distinguish them from non- or only slowly moving background
sources.

The first epoch observations of HD\,65216 and HD\,101930 were both obtained in June 2005 with
follow-up second epoch imaging one year later in June 2006. In order to limit saturation effects of
the detector due to the bright planet host stars, we always used the shortest available integration
time (1.2\,s), and 50 of these 1.2\,s integrations were averaged to one image. The standard dither
(jitter) technique was applied to subtract the bright infrared sky background. In total 10 images
were taken at 10 different dither positions, resulting in a total integration time of 10\,min. The
data reduction, i.e. background estimation and subtraction, flat-fielding of all images as well as
the final shift and add procedure was done with the ESO data reduction package \textsl{ECLIPSE}
(Devillard 2001).

Figure\,1 shows our second epoch SofI H-band images of HD\,65216 and HD\,101930, in which the
primaries are located in the center of each image.

\begin{figure}\resizebox{\hsize}{!}{\includegraphics{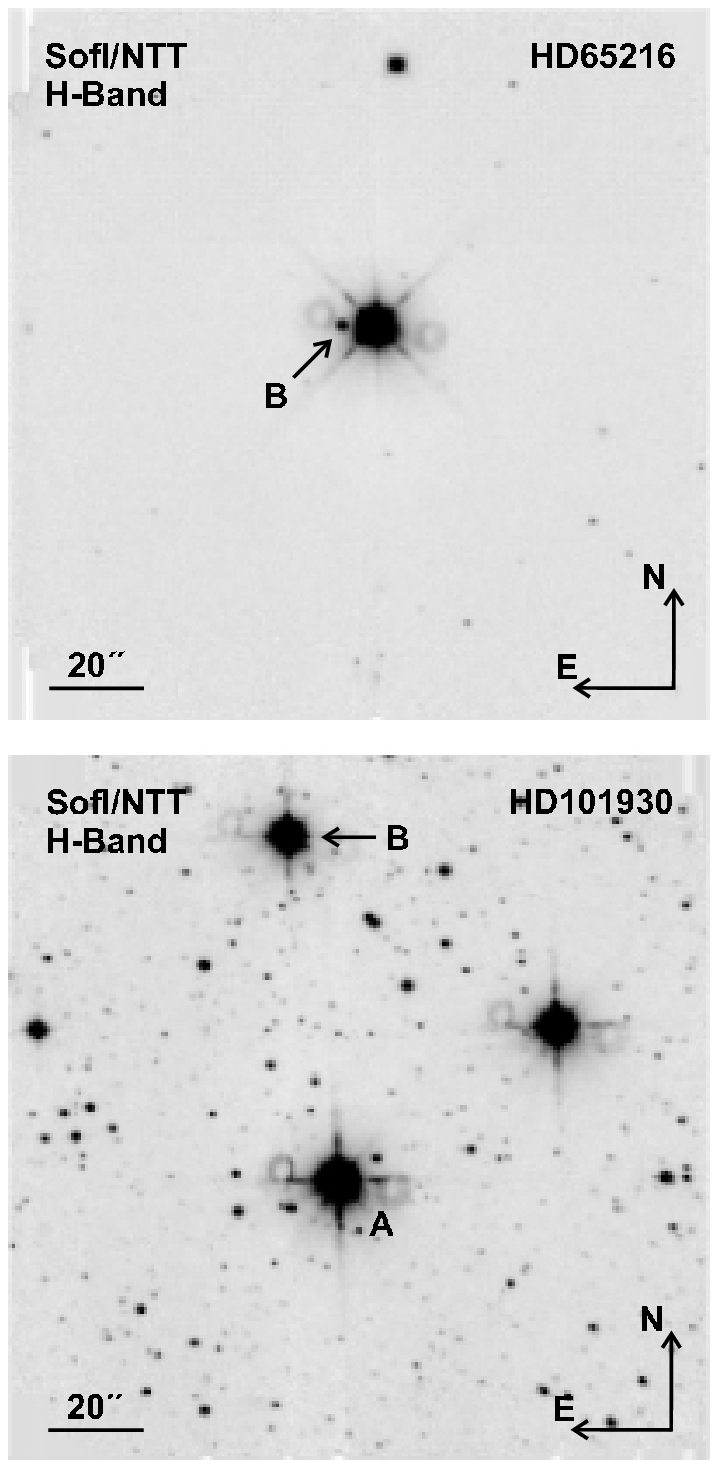}}
\caption{The SofI small field images of the planet host stars HD\,65216 (top) and HD\,101930
(bottom), taken in June 2006 in the H-band. The planet host stars are the bright stars in the
center of the images and the newly found co-moving companions are marked with a black arrow. Both
images show additional companion-candidates around the planet host stars down to a limiting
magnitude ($S/N=10$) of H$\sim$17\,mag, none of which are co-moving.} \label{pics}
\end{figure}

All images are astrometrically calibrated with sources detected in our SofI images and listed in
the 2MASS point source catalogue (Skrutskie et al. 2006). The astrometric calibration of both SofI
observing runs is summarized in Tab.\,\ref{data}.

\begin{table*}
\begin{center}
\caption{Pixel scale and detector orientation with their uncertainties for all SofI observing runs.
The detector is tilted by the given angle from north to west. Furthermore, the separations and
position angles of the detected companions HD\,65216\,B and HD\,101930\,B relative to their
primaries -- the exoplanet host stars HD\,65216\,A and HD\,101930\,A -- as well as their H-band
photometry are listed, as measured in all SofI observing epochs.}
\begin{tabular}{lcccc}
\hline
\textbf{Calibration \&}   & epoch       & pixel scale           & detector orientation & average seeing\\
\textbf{Seeing}        & date        & $[$arcsec/pixel$]$    & $[$$^{\circ}$$]$     & $[$arcsec$]$\\
\hline
                       & SofI 06/05  & 0.14330$\pm$0.00020   & 90.040$\pm$0.048 & 1.2\\
                       & SofI 06/06  & 0.14348$\pm$0.00016   & 90.017$\pm$0.049 & 1.0\\
\hline\hline
\textbf{Astrometry \&}    & epoch       & separation            & position angle        & H magnitude\\
\textbf{Photometry}    & date        & $[$arcsec$]$          & $[$$^{\circ}$$]$      & $[$mag$]$\\
\hline
HD\,65216\,B           & SofI 06/05  & \,\,\,7.097$\pm$0.030 & 89.68$\pm$0.25        & 12.674$\pm$0.063\\
                       & SofI 06/06  & \,\,\,7.144$\pm$0.028 & 89.39$\pm$0.23        & 12.685$\pm$0.063\\
\hline
HD\,101930\,B          & 2MASS 01/00 & 73.079$\pm$0.085      & \,\,\,8.28$\pm$0.07   & \,\,\,7.291$\pm$0.049\\
                       & \,\,\,\,\,\,\,\,\,SofI 06/05  & 73.033$\pm$0.122      & \,\,\,8.336$\pm$0.107 & ---\\
                       & \,\,\,\,\,\,\,\,\,SofI 06/06  & 73.120$\pm$0.102      & \,\,\,8.327$\pm$0.094 & ---\\
\hline
\end{tabular}
\end{center}
\label{data}
\end{table*}

\begin{figure}\resizebox{\hsize}{!}{\includegraphics{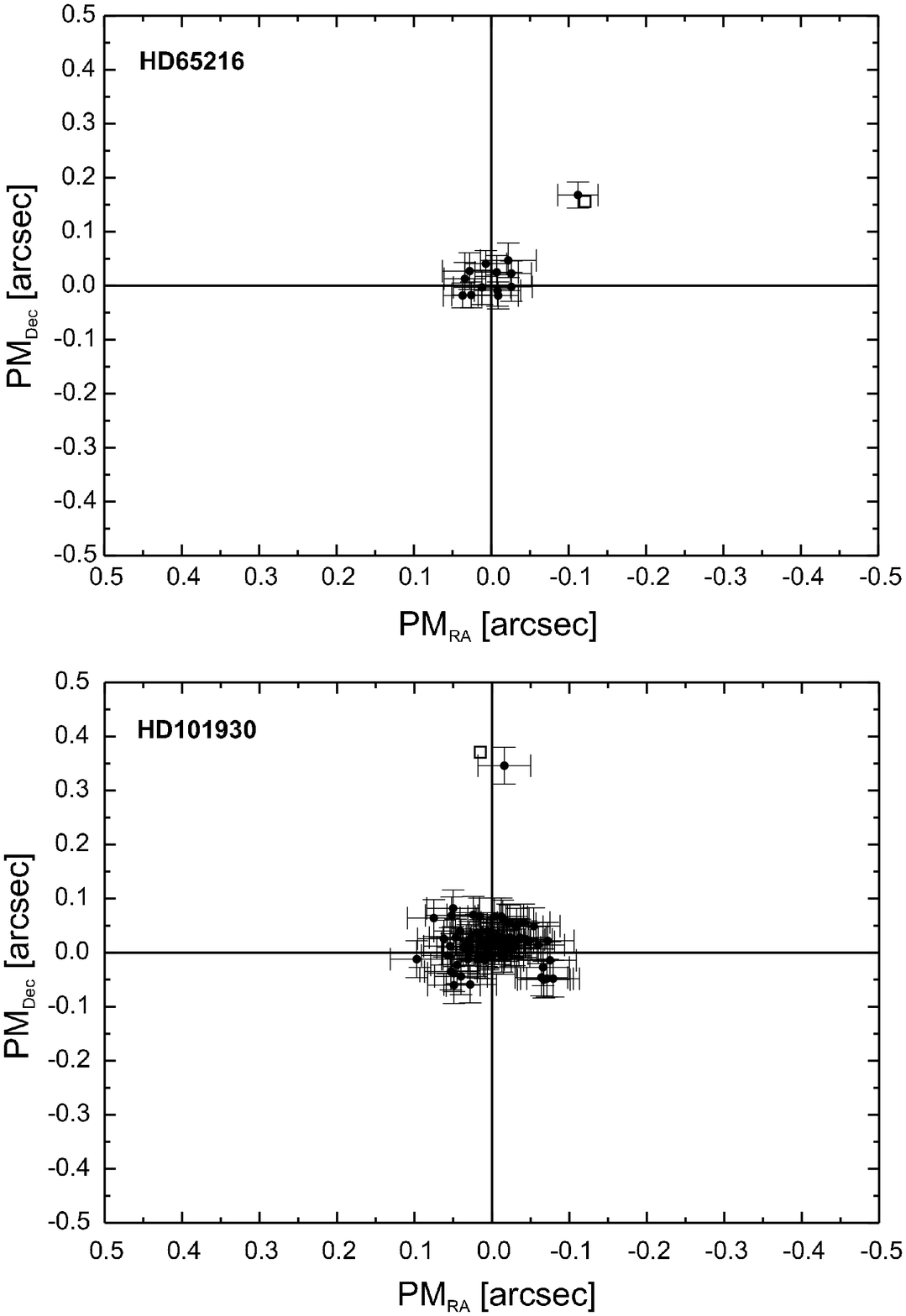}}
\caption{The measured proper motions of all detected ($S/N>10$) companion-candidates around the
planet host stars HD\,65216 (top) and HD\,101930 (bottom) between the SofI observing runs in June
2005 and June 2006. Most of the detected candidates exhibit only small or even negligible proper
motions, i.e. these sources are not companions of the fast moving planet host stars. In contrast,
two sources clearly share the proper motions of the planet host stars, which are indicated as
boxes. The position of these two co-moving companions HD\,65216\,B and HD\,101930\,B are indicated
in Fig.\,\ref{pics} and the SofI astrometry and photometry is summarized in Tab.\,\ref{data}.}
\label{astro}
\end{figure}

By comparing our first epoch H-band images with the second epoch images we determined the proper
motion of all objects ($S/N>10$) detected around the planet host stars within the SofI field of
view. As illustrated in Fig.\,\ref{astro}, most of the detected objects have a small or even
negligible proper motion. However, we have identified two sources which share the proper motion of
the planet host stars, indicated as boxes in Fig.\,\ref{astro}. Thus, these sources are co-moving
companions of the planet host stars and will be denoted HD\,65216\,B and HD\,101930\,B from hereon.

We measured the H-band photometry and relative astrometry of both detected co-moving companions in
our SofI H-band images. The results are summarized in Tab.\,\ref{data}. HD\,101930\,B is already
too bright and exceeds the linearity limit of the SofI detector, hence we cannot give an accurate
H-band photometry of this object. However, this object is well detected in 2MASS and accurate
photometry (photometric quality flag: AAA) of this companion is listed in the 2MASS point source
catalogue ($J=7.940\pm0.026$, $H=7.291\pm0.049$, $K_{S}=7.107\pm0.024$), see Skrutskie et al.
(2006). We used the 2MASS astrometry and derived the separation and position angle of HD\,101930\,B
relative to the planet host star at the epoch of the 2MASS observations (see Tab.\,\ref{data}).

HD\,101930\,B is also listed in the UCAC2 catalogue (Zacharias et al. 2004), in the USNO-B1.0
(Monet et al. 2003), as well as by Kharchenko (2001). The UCAC2
($\mu_{\alpha}cos(\delta)=25.6\pm3.3$mas/yr, $\mu_{\delta}=349.2\pm3.0$\,mas/yr) and USNO entry
($\mu_{\alpha}cos(\delta)=24$\,mas/yr, $\mu_{\delta}=348$\,mas/yr), as well as the astrometry of
HD\,101930\,B given by Kharchenko (2001) ($\mu_{\alpha}cos(\delta)=24.99\pm4.02$\,mas/yr,
$\mu_{\delta}=351.55\pm2.36$\,mas/yr) are fully consistent with our conclusion that this object and
the planet host star HD\,101930\,A form a common proper motion pair. In addition, Kharchenko (2001)
also lists the optical magnitude of HD\,101930\,B ($V=10.605\pm0.063$\,mag).

\begin{figure}\resizebox{\hsize}{!}{\includegraphics{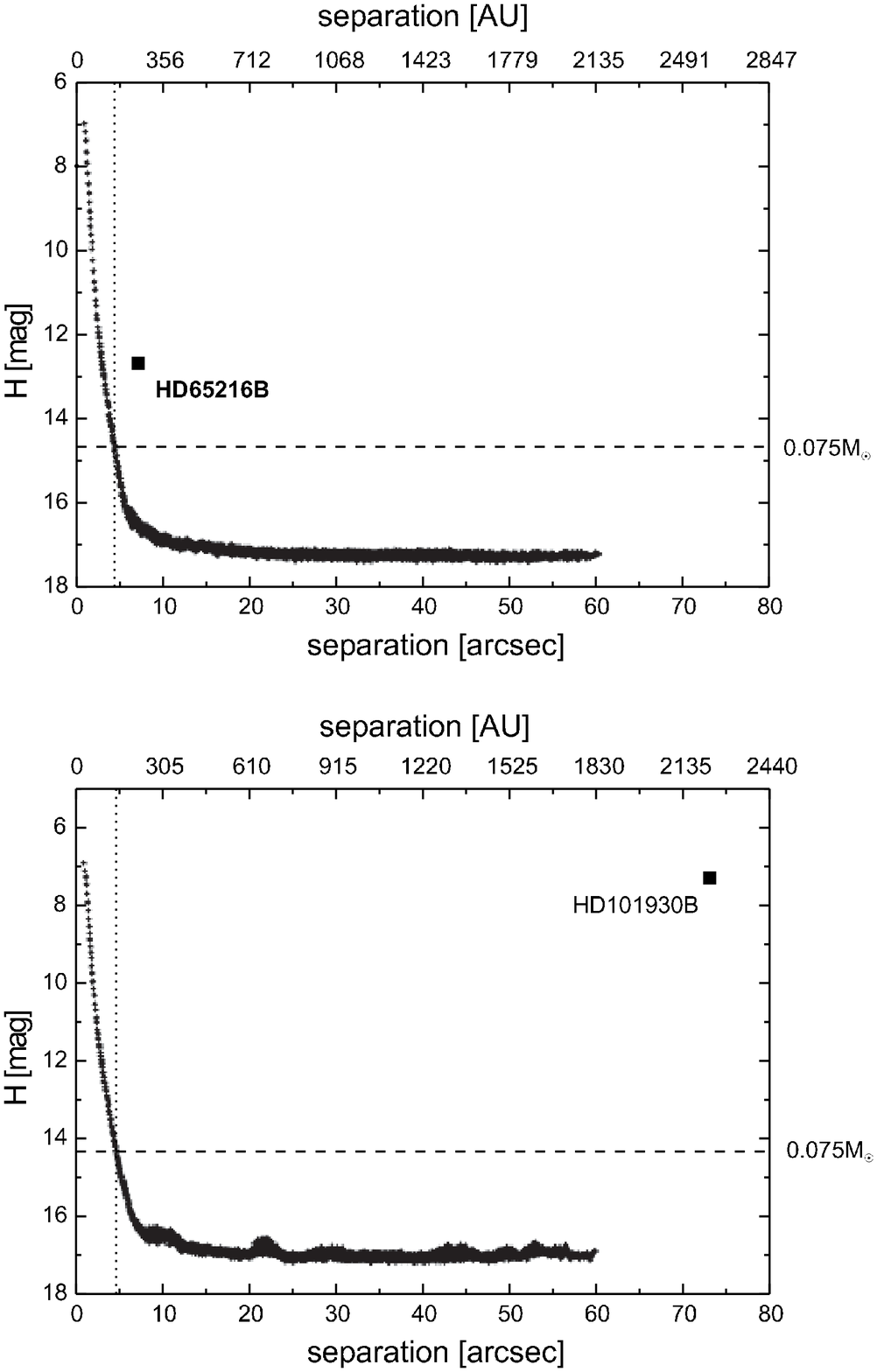}}
\caption{The detection limit ($S/N=10$) of our SofI H-band imaging of the exoplanet host stars
HD\,65216 (top) and HD\,101930 (bottom) plotted for a range of angular separations and projected
separations. The detected co-moving companions HD\,65216\,B and HD\,101930\,B are indicated as
black squares. The expected H-band magnitude of objects at the stellar-substellar border
(0.075\,$M_{\odot}$), is illustrated with a horizontal dashed line. All stellar companions can be
detected at angular separations larger than $\sim$5\,arcsec (see dotted vertical line) up to the
edge of the SofI field of view imaged twice in both observing epochs.} \label{limits}
\end{figure}

The detection limit of our SofI H-band images of both planet host stars is illustrated in
Fig.\,\ref{limits}. We obtain similar limits for both planet host stars. In the background limited
region beyond $\sim$\,15\,arcsec ($\sim$530\,AU of projected separation in case of HD\,65216 and
$\sim$460\,AU in case of HD\,101930, respectively), a limiting magnitude of $H\sim$17\,mag is
achieved. This allows the detection of substellar companions with masses down to
$\sim$65\,$M_{Jup}$, if we assume a system age of 5\,Gyr for both planet host star. This limit has
been derived using the Baraffe et al. (2003) evolutionary models. All stellar companions of the
planet host stars are detectable at angular separations larger than $\sim$5\,arcsec ($\sim$180\,AU
of projected separation in case of HD\,65216\,A and $\sim$150\,AU in case of HD\,101930\,A,
respectively) up to the edge of the SofI field of view imaged twice in both observing epochs, which
is $\sim 67$\,arcsec (2400\,AU of projected separation) around HD\,65216\,A and $\sim$59\,arcsec
(1800\,AU of projected separation) around HD\,101930\,A. Beside HD\,65216\,B and HD\,101930\,B no
further co-moving companions could be found within the SofI field of view around both planet host
stars.

In the case that both detected co-moving companions are low-mass stars located at the respective
distance of their primaries, we can approximate their masses from evolutionary stellar models.

In the case of HD\,65216\,B, only its apparent H-band magnitude is known, which is on average
$H=12.680\pm0.045$\,mag. The distance modulus $E=2.757\pm0.053$\,mag of the planet host star was
derived from the Hipparcos parallax. With the distance modulus and the apparent magnitude of
HD\,65216\,B we derive its absolute magnitude to be $M_{H}=9.923\pm0.069$\,mag. For an assumed age
of 5\,Gyr for the HD\,65216\,AB system, we can derive the mass of HD\,65216\,B from evolutionary
models of low-mass stars and substellar objects. According to the Baraffe et al. (1998) models, the
absolute magnitude of HD\,65216\,B is consistent with a low-mass star of
$0.094\pm0.002\,M_{\odot}$.

In the case of HD\,101930\,B, optical as well as infrared photometry is available. The Hipparcos
parallax of the planet host star HD\,101930\,A yields a distance modulus $E=2.421\pm0.064$\,mag.
With the given apparent H-band photometry of the companion we obtain its absolute magnitude
$M_{H}=4.870\pm0.080$\,mag. If we assume again 5\,Gyr as the age of the HD\,101930\,AB system we
obtain $0.666\pm0.013\,M_{\odot}$ as the mass of HD\,101930\,B using the Baraffe et al. (1998)
evolutionary models as well as the derived absolute H-band photometry of the companion.

The mass approximations described above have to be confirmed with spectroscopy, i.e. it has to be
shown that the detected companions are indeed low-mass stellar objects. In particular companions as
faint as HD\,65216\,B in the near infrared could also be white dwarfs. However, the spectra of
these degenerated objects are clearly different to those of low-mass stars. Spectroscopy finally
determines the true nature of the companions, which will also confirm their companionship to the
planet host stars.

\section{Infrared spectroscopy}

\subsection{HD\,65216\,B and HD\,101930\,B}

We obtained infrared H- and K-band spectroscopy of HD\,65216\,B in December 2006 with VLT-ISAAC,
the infrared imager and spectrograph on UT1 (Antu) at Paranal observatory. We used ISAAC's
low-resolution spectroscopy mode SWS1-LR with the 1\,arcsec slit, providing a resolving power of
500 in H-band, and 450 in K-band, with a dispersion of 4.8\,\AA\,\,per pixel in H-band, and
7.2\,\AA\,\,per pixel in K-band.

In the H-band we took 12 frames each the average of two 30\,s integrations, i.e. 12\,min of total
integration time. In order to remove the high infrared background, the telescope was always nodded
50\,arcsec between two positions along the slit. In addition we applied a 5\,arcsec dither to the
nodding positions in order to avoid that the spectrum always falls on the same pixels on the ISAAC
detector. We took 14 frames in K-band, each the average of two 30\,s integrations, i.e. 14\,min of
total integration time. For wavelength calibration we took spectra of a Xenon lamp. All frames were
flat-fielded and the spectra are extracted from the individual frames, wavelength calibrated and
finally averaged using standard IRAF data reduction routines. Telluric features were removed,
dividing by a spectrum of the telluric standard star HIP\,52670 (B3V) which was taken between the
H- and K-band spectroscopy of HD\,65216\,B. Thus, the airmass difference between science and
calibration spectra was minimized to less than 0.1. The spectral response function of ISAAC was
determined, using the spectra of the telluric standard and flux-calibrated B3V spectra from the
spectral library of Pickles et al. (1998).

The spectra of the wide companion HD\,101930\,B were taken at the beginning of February 2007 with
SofI in its low-resolution spectroscopy mode. We used the grism RED in combination with a 1\,arcsec
slit, providing a resolving power of 588 in H- and K-band with a dispersion of 10.22\,\AA\,\,per
pixel. Six frames, each the average of 20 times 3\,s integrations, yield a total integration time
of 6\,min. The NTT was nodded along the direction of the slit between two positions, separated from
each other by 37\,arcsec. In addition a random dither of 2\,arcsec was applied around both nodding
positions. A spectrum of a Xenon lamp was taken for wavelength calibration. All frames were
flat-fielded and the individual spectra were extracted, wavelength calibrated and finally averaged
using again standard IRAF routines. We took a spectrum of the telluric standard HIP\,54930 (B1V)
immediately after HD\,101930\,B to minimize airmass difference between the science and calibration
observations to less than 0.1. Telluric features in the spectra of HD\,101930\,B were removed with
the spectra of the telluric standard, which was again used together with the flux-calibrated
spectra from the spectral library of Pickles et al. (1998) to determine the spectral response
function of SofI.

\begin{figure}\resizebox{\hsize}{!}{\includegraphics{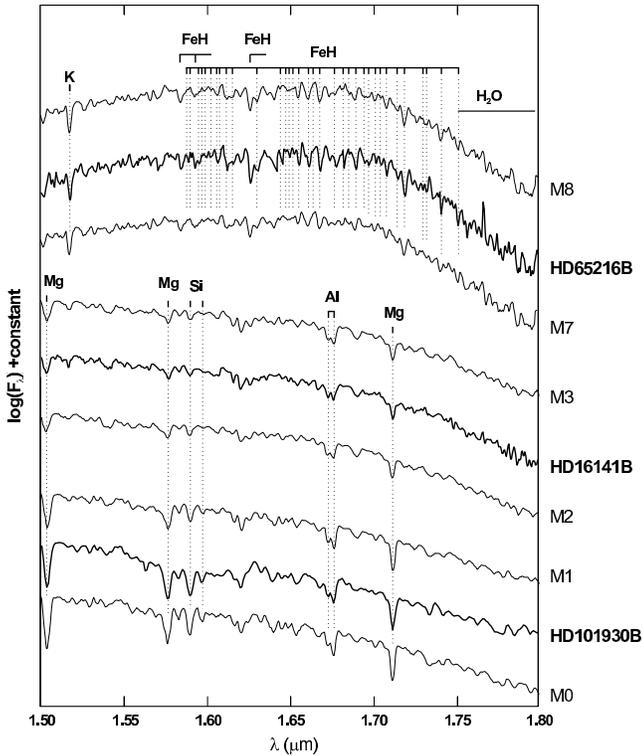}}
\caption{The ISAAC H-band spectra of HD\,65216\,B and HD\,16141\,B and the SofI H-band spectrum of
HD\,101930\,B together with comparison spectra of dwarfs from the IRTF Spectral Library (Cushing et
al. 2005). The most prominent spectral atomic and molecular features are indicated. The shape of
the continuum as well as all detected spectral features of the spectrum of HD\,101930\,B are
consistent with the comparison H-band spectra of M0 and M1 dwarfs. The H-band spectrum of
HD\,16141\,B is most comparable with those of M2 to M3 dwarfs and the spectrum of HD\,65216\,B with
those of M7 to M8 dwarfs.} \label{hspec}
\end{figure}

\begin{figure}\resizebox{\hsize}{!}{\includegraphics{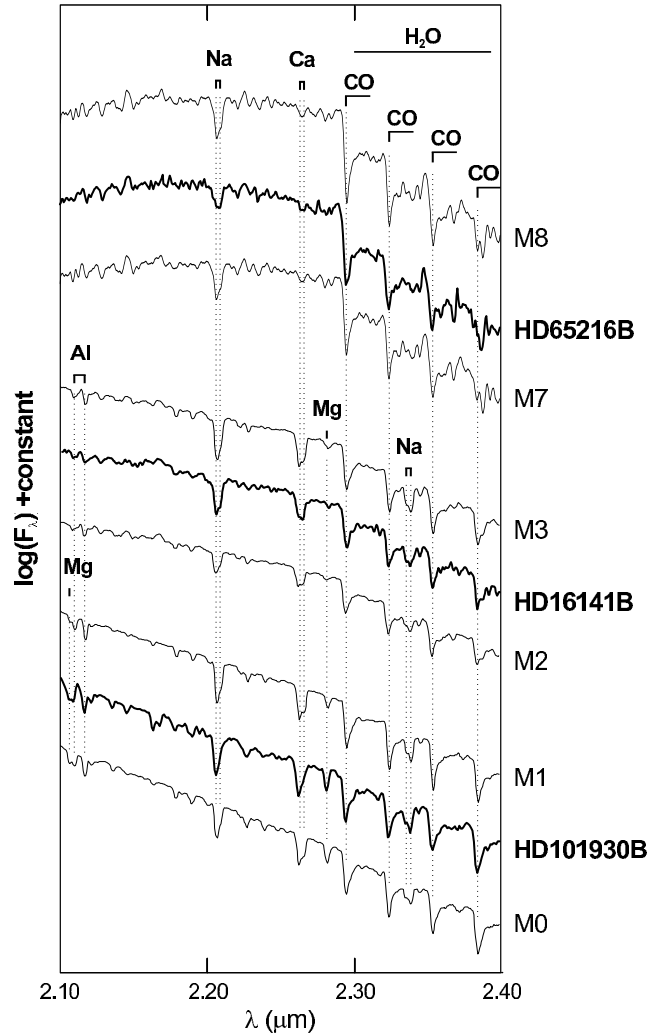}}
\caption{The ISAAC K-band spectra of HD\,65216\,B and HD\,16141\,B and the SofI K-band spectrum of
HD\,101930\,B together with comparison spectra of dwarfs from the IRTF spectral library (Cushing et
al. 2005). The most prominent spectral atomic and molecular features are indicated. The shape of
the continuum as well as all detected spectral features of the spectrum of HD\,101930\,B are
consistent with the comparison K-band spectra of M0 and M1 dwarfs. The K-band spectrum of
HD\,16141\,B is most comparable with those of M2 to M3 dwarfs and the spectrum of HD\,65216\,B with
those of M7 to M8 dwarfs.} \label{kspec}
\end{figure}

The reduced and flux-calibrated H- and K-band spectra of HD\,65216\,B and HD\,101930\,B are shown
in Fig.\,\ref{hspec} and Fig.\,\ref{kspec}. We compare the H- and K-band spectra of both companions
with template spectra from the IRTF spectral library (Cushing et al. 2005). To compare our SofI and
ISAAC spectra with these templates, the template spectra where smoothed to the same resolution
($\Delta \lambda / \lambda = 1/500$) in the H- and K-band.

The continua of the H- and K-band spectra of HD\,65216\,B as well as all detected atomic and
molecular features are mostly consistent with comparison spectra of spectral type M7V and M8V. In
the H-band the most prominent features in the spectrum of HD\,65216\,B are those of Potassium at
1.517\,$\mu$m, the line series of FeH (strongest feature at 1.625\,$\mu$m) as well as the flux
depression induced by H$_{\rm 2}$O at wavelength longer than 1.75\,$\mu$m. The overall shape of the
continuum in the H-band is due to collision induced absorption (CIA) by H$_2$. In the K-band atomic
features of Sodium (Na doublet at 2.208\,$\mu$m) and Calcium (doublet at 2.265\,$\mu$m) are
detected. However, the strongest features in the K-band spectrum are those of the CO series
extending from the band head at 2.294\,$\mu$m to longer wavelength. At wavelength longer than
2.3\,$\mu$m we see again a flux depression induced by H$_{\rm 2}$O, typical for a dwarf of spectral
type M7 to M8.

In the H-band spectrum of HD\,101930\,B the most prominent spectral features are the atomic
absorbtion lines of Magnesium at 1.504\,$\mu$m, 1.576\,$\mu$m, 1.711\,$\mu$m, Silicium at
1.589\,$\mu$m and Aluminium at 1.674\,$\mu$m. The detected absorption features as well as the
continuum of the H-Band spectrum of the companion are consistent with those of comparison spectra
of M0 to M1 dwarfs.

The same holds for the K-band spectrum of HD\,101930\,B where we find the strong absorption line
doublets of Sodium at 2.208\,$\mu$m and Calcium 2.265\,$\mu$m as well as the CO absorption band at
wavelength longer than 2.294\,$\mu$m. In addition the weaker absorption lines of the Sodium doublet
at 2.337\,$\mu$m, the absorption lines of Magnesium at 2.107\,$\mu$m and 2.282\,$\mu$m, as well as
Aluminium at 2.117\,$\mu$m are detected.

The determined spectral types of HD\,65216\,B and HD\,101930\,B are fully consistent with the
apparent magnitudes of both co-moving companions if one assumes that both objects are located at
the distances of the planet host stars, as it is expected for real companions. Hence, the
companionship of HD\,65216\,B and HD\,101930\,B revealed first by astrometry is finally confirmed
by photometry and spectroscopy.

\subsection{HD\,16141\,B}

We reported a co-moving companion located $\sim$6\,arcsec ($\sim$220\,AU of projected separation)
south of the planet host star HD\,16141 in Mugrauer et al. (2005). We determined the H-band
photometry of the companion to be $H=10.062\pm0.049$\,mag. With the known distance of the planet
host star (Hipparcos parallax $\pi=27.85\pm1.39$\,mas) this yields $M_{H}=7.286\pm0.119 $\,mag as
absolute magnitude of the companion. By assuming that the companion is a low-mass stellar object
the mass of the companion was determined using again the stellar evolutionary models from Baraffe
et al. (1998). For an assumed age of the companion of 5\,Gyr we obtain
$0.286\pm0.017$\,$M_{\odot}$. The expected color of such an object is also given by the models to
$V-K=4.52\pm0.06$\,mag. According to the spectral type -- color relation from Kenyon \& Hartmann
(1995) the derived color is consistent with a M3 dwarf.

In order to confirm this result and to finally to determine the true nature of HD\,16141\,B we
obtained follow-up spectra in October 2006 with ISAAC, using the same setup as for HD\,65216\,B.
For correction of telluric features and for the flux-calibration we observed the telluric standard
Hip\,15188 (B3V). The reduced H- and K-band spectra of HD\,16141\,B are shown in Fig.\,\ref{hspec}
and Fig.\,\ref{kspec}.

The continua of the H- and K-band spectra as well as all detected atomic and molecular absorption
features are consistent with a spectral type in the range M2V to M3V, i.e. HD\,16141\,B is a low
mass stellar object as it was already expected from its photometry, adopting the distance of the
planet host star for its companion.

\section{Discussion}

As described in the last two sections we can confirm the companionship of HD\,65216\,B and
HD\,101930\,B detected first by astrometry as co-moving objects to the planet host stars
HD\,65216\,A and HD\,101930\,A, with photometry and spectroscopy. The photometry of both companions
is consistent with low-mass stellar objects located at the distances of the planet host stars,
which is finally confirmed with ISAAC and SofI spectroscopy.

HD\,101930\,B is separated from its primary by $\sim$73\,arcsec which corresponds to a projected
separation of 2229\,AU at the distance of the planet host star HD\,101930\,A. With the mass of the
planet host star (0.74\,$M_{\odot}$) and its stellar companion (0.666\,$M_{\odot}$) as well as with
the separation of the system we can approximate the longtime stable region for additional
companions as derived by Holman \& Wiegert (1999) for assumed circular companion orbits. By
assuming an orbital eccentricity $e=0.5$ for HD\,101930\,B we obtain $a_{c}$=270\,AU and
$a_{c}$=633\,AU for a circular ($e=0$) orbit, respectively. Thus, the planet of HD\,101930\,A
clearly resides well within the longtime stable region of its host star.

\subsection{Evidence for the binarity of HD\,65216\,B}

\begin{figure}\resizebox{\hsize}{!}{\includegraphics{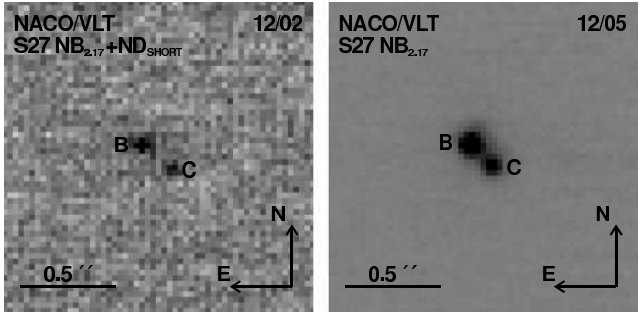}}
\caption{NACO images of HD\,65216\,B taken in the K-band through the narrow band filter
NB$_{\rm{2.17}}$. The co-moving companion is resolved in two objects B and C, separated by less
than 0.2\,arcsec.} \label{nacopics}
\end{figure}

HD\,65216\,B is separated from its primary by $\sim$7\,arcsec which corresponds to a projected
separation of 253\,AU at the distance of the planet host star HD\,65216\,A. Besides our SofI H-band
observations the HD\,65216 system was also observed with the adaptive optics system NACO at UT4
(Yepun) at Paranal observatory. We retrieved public data from the ESO archive obtained with the S27
camera (pixel scale: 27.15\,mas per pixel and 28\,arcsec$\times$28\,arcsec field of view) of NACO
in December 2002 and December 2005. In the first observing run 28 frames, each the average of two
15\,s integrations, were taken in dither mode through the narrow band filter NB$_{\rm{2.17}}$ and
the neutral density filter ND$_{\rm{SHORT}}$, i.e. 14\,min of total integration time. In the second
run again 28 frames were taken in dither mode, each the average of 60 integrations of 0.5\,s
through the narrow band filter NB$_{\rm{2.17}}$, i.e. again 14\,min of total integration time. The
images were flat-fielded and combined using the data reduction package \textsl{ECLIPSE} (Devillard
2001). At the position of HD\,65216\,B two objects are detected which are shown in
Fig.\,\ref{nacopics} for both NACO observing epochs.

The brighter source B is separated by only $189\pm11$\,mas from the fainter object C in the first
epoch NACO image from December 2002 (they could also be called Ba and Bb, but we prefer B and C).
In the second epoch NACO image both sources are separated by only $146\pm3$\,mas, i.e. the
separation significantly decreased by $\Delta sep=-43\pm12$\,mas. If we assume that the brighter
component B follows the proper motion of the planet host star but the faint source C is just a
non-moving background source, the expected change of separation should be much larger, namely
$\Delta sep=-423$\,mas (15.6 NACO S27-pixels) for the given epoch difference between both NACO
observation runs. This expected change of separation is about ten times larger than the one we
measured. Hence, we conclude that B and C are co-moving objects, i.e. the HD\,65216 system is a
hierarchical triple system composed of the planet host star HD\,65216\,A which is orbited by the
close binary system HD\,65216\,BC at a wide separation (253\,AU) and by a planet candidate at close
separation (1.37\,AU).

The measured change of the separation between the B and C components can be explained by orbital
motion of the B and C components around their common center of mass. In order to estimate this
orbital motion we have to determine first the masses of both components.

In the second epoch narrow band NACO images all three components of the HD\,65216 system are
detected within the linearity limit of the NACO detector. We determine a magnitude difference of
$\Delta \rm{NB_{\rm{2.17}}}(AB)=6.31\pm0.02$\,mag and $\Delta
\rm{NB_{\rm{2.17}}}(AC)=7.32\pm0.05$\,mag. With the given 2MASS $K_{S}$-band magnitude of
HD\,65216\,A we can approximate the $K_{S}$-band magnitudes of HD\,65216\,B and C and obtain
$K_{S}(B)=12.64\pm0.03$\,mag and $K_{S}(C)=13.65\pm0.06$\,mag. With the known distance of the
planet host star this finally yields the absolute $K_{S}$-band magnitudes of the two components
$M_{Ks}(B)=9.88\pm0.06$\,mag and $M_{K_{S}}(C)=10.89\pm0.08$\,mag.

According to the spectral type -- magnitude relation from Cruz et al. (2003) these values are
consistent with low-mass objects of spectral type M7 to M8 for HD\,65216\,B and L2 to L3 for
HD\,65216\,C. The absolute H-band magnitude of a M7 to M8 dwarf is $M_{H}=10.3\pm0.3$\,mag and
$M_{H}=11.5\pm0.2$\,mag for objects of spectral types between L2 to L3. The same result is obtained
by an independent analysis of the same public NACO datasets done by A. Eggenberger (private
communication), see Eggenberger et al. (2007, in preparation). Hence, the total magnitude of the
unresolved HD\,65216\,BC system should be $M_{H}=9.99\pm0.24$\,mag, which corresponds to an
apparent magnitude of $H=12.75\pm0.25$\,mag at the distance of the planet host star. It is worth to
mention that the SofI H-band photometry of the unresolved pair HD\,65216\,BC
$H=12.680\pm0.045$\,mag is fully consistent with the derived magnitude estimate.

If we assume a system age of 5\,Gyr we can use again the Baraffe et al. (1998) models to derive the
masses of both components. We obtain $0.089\pm0.001\,M_{\odot}$ for HD\,65216\,B and
$0.078\pm0.001\,M_{\odot}$ for HD\,65216\,C, respectively (since the system is older according to
its metallicity, the true masses may be slightly larger). Hence the total mass of HD\,65216\,BC is
0.167$\,M_{\odot}$. The average separation of both components is 167\,mas, which corresponds to a
projected separation of 6\,AU at the distance of the planet host star. We can use this separation
as estimate for the semi-major axis of the system and finally derive its orbital period using
Kepler's third law, which yields 36 years. The maximal expected change of separation between both
components is 19\,mas/yr. Thus, for an epoch difference of three years the expected maximal change
of separation due to orbital motion should be smaller than 57\,mas. Indeed, the measured change of
separation between HD\,65216\,B and C ($\Delta sep=-43\pm12$\,mas) can be explained with orbital
motion of both objects around their barycenter. This relatively short orbital motion is important,
because it will yield directly determined masses from astrometry and Kepler's third law within a
few decades, which are very rare at such late spectral types and low masses, hence will be very
useful for calibrating models and to constrain the mass-luminosity relation at the low-mass end.

Finally, we can determine the radius of the longterm stable region around the planet host star,
using again the approximation of Holman \& Wiegert (1999) and the derived masses of HD\,65216\,B
and C (total mass 0.167$\,M_{\odot}$), the mass of the planet host star (0.94$\,M_{\odot}$) as well
as the projected separation between the star and its binary companion (253\,AU). We obtain
$a_{c}=42$\,AU for an assumed eccentric orbit of HD\,65216\,BC with $e=0.5$ and $a_{c}=103\,AU$ for
a circular orbit, respectively.

\section*{Acknowledgments}

We would like to thank the technical staff of the ESO NTT for all help and assistance in carrying
out the observations. Furthermore, we would like to thank T. Mazeh for his participation in our
long-term project to study the multiplicity of exoplanet host stars and for all his helpful
suggestions. We made use of the 2MASS public data releases, as well as the Simbad database operated
at the Observatoire Strasbourg.

\label{lastpage}

\end{document}